\documentclass{article}
\usepackage{ltwol2e}

\usepackage{epsfig}

\begin{document}

\title{THE DUALIZED STANDARD MODEL AND ITS APPLICATIONS}

\author{CHAN Hong-Mo}

\address{Rutherford Appleton Laboratory,
  Chilton, Didcot, Oxon, OX11 0QX, United Kingdom
  \\E-mail: chanhm\,@\,v2.rl.ac.uk}

\author{Jos\'e BORDES}

\address{Dept. Fisica Teorica, Univ. de Valencia,
  c. Dr. Moliner 50, E-46100 Burjassot (Valencia), Spain\\
  E-mail: jose.M.bordes\,@\,uv.es}

\author{TSOU Sheung Tsun}

\address{Mathematical Institute, University of Oxford,
  24-29 St. Giles', Oxford, OX1 3LB, United Kingdom
  \\E-mail: tsou\,@\,maths.ox.ac.uk}

\twocolumn[\maketitle\abstracts{ The Dualized Standard Model offers a
natural explanation for Higgs fields and 3 generations of fermions
plus a perturbative method for calculating SM parameters.  By
adjusting only 3 parameters, 14 quark and lepton masses and mixing
parameters (including $\nu$ oscillations) are calculated with general
good success.  Further predictions are obtained in post-GZK air
showers and FCNC decays.}]

In this article, we summarize some work which has occupied us for some
years.  The material has been summarized in 5 papers submitted to this 
conference (Papers 607, 610, 611, 613, 636), and this talk is but a 
summary of these summaries.  

The Dualized Standard Model \cite{Chantsou1} is a scheme which aims to 
answer some of the questions left open by the Standard Model (such as 
why Higgs fields or fermion generations should exist) and to explain the 
values of some the Standard Model's many parameters (such as fermion 
masses and mixing angles).  In contrast to most schemes with similar 
aims, the DSM remains entirely within the Standard Model framework, 
introducing neither supersymmetry nor higher space-time dimensions.  
That it is able to derive results beyond the Stardard Model while 
remaining within its framework is by exploiting a generalization of 
electric-magnetic duality to nonabelian Yang-Mills theory found a couple 
of years ago \cite{Chanftsou1}.

The concept of duality is best explained by recalling the well-known
example in electromagnetism.  There a dual transform (the Hodge star) 
is defined:
${}^*\!F_{\mu\nu} = -(1/2) \epsilon_{\mu\nu\rho\sigma} F^{\rho\sigma}$,
where for both the Maxwell field $F_{\mu\nu}$ and its dual 
${}^*\!F_{\mu\nu}$
potentials $A_\mu$ and $\tilde{A}_\mu$ exist,
so that the theory is invariant under both $A_\mu$ and $\tilde{A}_\mu$ 
gauge transformations.
The theory has thus in all a $U(1) \times \tilde{U}(1)$ gauge symmetry
with $U(1)$ corresponding to electricity and $\tilde{U}(1)$ to magnetism.
Magnetic charges are monopoles in $U(1)$, while electric charges are 
monopoles of $\tilde{U}(1)$.

The same statements do not hold for nonabelian Yang-Mills theory under
the dual transform \cite{Guyang} (star).  However, it was
shown \cite{Chanftsou1} that there exists a generalized dual transform
for which similar results apply.   Its exact form,
in the language of loop space \cite{Polyakov,Chantsou2},
need not here bother us.  What matters, however, is that given this
generalized transform, a potential can again be defined for both the 
Yang-Mills field and its dual, and that the theory is invariant under
both the gauge transformations:
\begin{equation}
A_\mu \to A_\mu +\partial_\mu \Lambda + ig [\Lambda, A_\mu],
\ 
\tilde{A}_\mu \to \tilde{A}_\mu +\partial_\mu \tilde{\Lambda}
   + i \tilde{g} [\tilde{\Lambda}, \tilde{A}_\mu],
\label{Attransf}
\end{equation}
with $g, \tilde{g}$ satisfying the (generalized) Dirac quantization 
condition \cite{Chantsou3}: $g \tilde{g} = 4 \pi$.
As a result, the theory has in all the gauge symmetry $SU(N) \times
\widetilde{SU}(N)$ with $SU(N)$ corresponding to (electric) `colour'
and $\widetilde{SU}(N)$ to (magnetic) `dual colour'.  And again, dual 
colour charges are monopoles in $SU(N)$, while colour charges appear 
as monopoles in $\widetilde{SU}(N)$ \cite{Chanftsou2}.

Applied to colour in the Standard Model, this nonabelian 
duality \cite{Chanftsou1} gives two new interesting features.  First, dual 
to the $SU(3)$ symmetry of colour, the theory possesses also an 
$\widetilde{SU}(3)$ symmetry of dual colour.  Then, by a well-known 
result of 't~Hooft \cite{thooft}, since colour is confined, it follows
that this $\widetilde{SU}(3)$ of dual colour has to be broken via a
Higgs mechanism\footnote{It has been shown \cite{Chantsou3} the 
duality introduced \cite{Chanftsou1} indeed
satisfies the commutation relations of the order-disorder parameters
used by 't~Hooft to define his duality.}.
Hence, the theory already contains within itself a broken 3-fold gauge
symmetry which could play the role of the `horizontal' symmetry 
wanted to explain the existence in nature of the 3 
fermion generations.  Second, in the generalized dual 
transform \cite{Chanftsou1}, the frame vectors (`dreibeins') in the 
gauge group
take on a dynamical role \cite{Chanftsou2} which suggests that they 
be promoted to physical fields.  If so, then they possess
the properties that one wants for Higgs fields for symmetry
breaking (as in electroweak theory): space-time scalars belonging
to the fundamental representation having classical
values (vev's) with finite lengths.  

The basis of the Dualized Standard Model is just in making this bold
assumption of identifying the dual colour $\widetilde{SU}(3)$ as
the `horizontal' generation symmetry and of the frame vectors in it
as the Higgs fields for its breaking.
We note that according to nonabelian duality \cite{Chanftsou1},
the niches already exist in the original theory in the form of the dual
symmetry and the `dreibeins'.  One could 
thus claim that the DSM offers an explanation for the existence 
both of exactly 3 fermion generations and of Higgs fields necessary 
for breaking this generation symmetry.

This identification further suggests the manner in which the symmetry ought
to be broken.  As a result, the fermion mass matrix at tree-level takes
the form \cite{Chantsou1}:
\begin{equation}
m = m_T \left( \begin{array}{c} x \\ y \\ z \end{array} \right) (x,y,z),
\label{massmat0}
\end{equation}
where $m_T$ is a normalization factor depending on the fermion-type $T$,
namely whether $U$- or $D$-type quarks, charged leptons ($L$) or neutrinos
($N$), and $x, y, z$ are vacuum expectation values of Higgs fields
(normalized for convenience: $x^2 + y^2 + z^2 = 1$),
which are independent of the fermion-type $T$.  Because $m$ is
factorizable it has only one nonzero
eigenvalue so that at tree-level there is only one massive generation
(fermion mass hierarchy).  Further, because $(x, y, z)$ 
is independent of the fermion-type, the state vectors of,
say, the $U$- and $D$-type quarks in generation space have the
same orientation, so that the CKM matrix is the unit matrix.  These are
already not a bad first approximation to the experimental situation.

One can go further, however.  With loop corrections, it is seen that
the mass matrix $m'$ remains factorizable~\cite{Chantsou1}, with the
same form as (\ref{massmat0}), but
the vector $(x', y', z')$, in which the relevant information of
$m'$ is encoded, now rotates with the energy scale, tracing out 
a trajectory on the unit sphere.  Hence, the
lower generation fermions acquire small finite masses via `leakage'
from the highest generation.  Furthermore, the vector $(x', y', z')$
depends now on the fermion-type,  
giving rise to a nontrivial CKM matrix.  The result is a perturbative
method for calculating fermion mass and mixing parameters.

In a 1-loop calculation 
\cite{Bordesetal1,Bordesetal2} it is found that out of 
the many diagrams  only the Higgs loop
diagram dominates, involving thus only a few adjustable parameters.
The present score is as follows.  By adjusting 3 parameters, namely a
Yukawa coupling strength $\rho$ and the 2 ratios between the Higgs vev's
$x, y, z$, one has calculated the following 14 of the `fundamental'
SM parameters:\\
\hspace*{.5cm}$\bullet$ the 3 parameters of the quark CKM matrix
$|V_{rs}|$,\\
\hspace*{.5cm}$\bullet$ the 3 parameters of the lepton CKM matrix
$|U_{rs}|$,\\
\hspace*{.5cm}$\bullet$ $m_c, m_s, m_\mu, m_u, m_d, m_e$,\\
\hspace*{.5cm}$\bullet$ the masses $m_{\nu_1}$ of the lightest and $B$
of the right-\\
\hspace*{.8cm}handed neutrinos,\\
there being no $CP$-violation at 1-loop order.  

First, for the quark CKM matrix $|V_{rs}|$, where $r = u, c, t$ and $s =
d, s, b$, one obtains for a sample fit \cite{Chantsou4}:
\begin{equation}
|V_{rs}| =
\left( \begin{array}{ccc} 0.9752 & 0.2215 & 0.0048 \\
                          0.2210 & 0.9744 & 0.0401 \\
                          0.0136 & 0.0381 & 0.9992 \end{array} \right),
\label{calckmq}
\end{equation}
as compared with the experimental values \cite{databook}:
\begin{equation}
\left( \begin{array}{lll} 0.9745-0.9760 & 0.217-0.224 & 0.0018-0.0045 \\
                          0.217-0.224 & 0.9737-0.9753 & 0.036-0.042 \\
                          0.004-0.013 & 0.035-0.042 & 0.9991-0.9994
   \end{array} \right).
\label{expckmq}
\end{equation}
All the calculated values are seen to lie roughly within the experimental
bounds.

Second, for the lepton CKM matrix $|U_{rs}|$, one obtains with the
{\em same} 3 input parameters:
\begin{equation}
|U_{rs}| = \left( \begin{array}{ccc} 0.97 & 0.24 & 0.07 \\
                                     0.22 & 0.71 & 0.66 \\
                                     0.11 & 0.66 & 0.74 \end{array} \right),
\label{calckml}
\end{equation}
where $r = e, \mu, \tau$ and $s = 1, 2, 3$ label the physical states of
the neutrinos.    The empirical values of $|U_{rs}|$ for leptons are 
much less well-known.  Collecting all 
the information so far available from neutrino oscillation experiments, 
one arrives at the following tentative arrangement:
\begin{equation}
|U_{rs}| = \left( \begin{array}{ccc} \star & 0.4-0.7 & 0.0-0.15 \\
                                     \star & \star & 0.45-0.85 \\
                                     \star & \star & \star \end{array} \right).
\label{expckml}
\end{equation}
which is seen to be in very good agreement with the prediction (\ref{calckml}) 
for $U_{e3}$ and $U_{\mu3}$, but not for $U_{e2}$.  

Lastly, from the {\em same} calculation with the {\em same} 3
parameters, one obtains 
the fermion masses listed in Table \ref{Masses}.  The  
\begin{table}
\caption{Fermion Masses}
\vspace*{-.5cm}
\begin{eqnarray*}
\begin{array}{|c|c|c|}
\hline
& Calculation & Experiment \\
\hline
m_c & 1.327 {\rm GeV} & 1.0-1.6 {\rm GeV} \\
m_s & 0.173 {\rm GeV} & 100-300 {\rm MeV} \\
m_\mu & 0.106 {\rm GeV} & 105.7 {\rm MeV} \\
m_u & 235 {\rm MeV} & 2-8 {\rm MeV} \\
m_d & 17 {\rm MeV} & 5-15 {\rm MeV} \\
m_e & 7 {\rm MeV} & 0.511 {\rm MeV} \\
m_{\nu_1} & 10^{-15} {\rm eV} & < 10 {\rm eV} \\
B & 400 {\rm TeV} & ? \\
\hline 
\end{array}
\end{eqnarray*}
\vspace*{-.5cm}
\label{Masses}
\end{table}
second generation masses agree very well with experiment.  Those
of the  
lowest generation were obtained by extrapolation on
a logarithmic scale and should be regarded as reasonable if of roughly
the right magnitude.
As for the 2 neutrino
masses, the experimental bounds are so weak that there is essentially
no check.

In summary, out of the 14 quantities calculated, 8 are good 
($|V_{rs}|, |U_{\mu3}|, |U_{e3}|, m_c, m_s, m_\mu$), 2 are 
reasonable ($m_d, m_e$), 2 are unsatisfactory ($|U_{e2}|, m_u$), and 2 
are untested, which is not a bad score for a first-order calculation 
with only 3 parameters.

\begin{figure}[htb]
\vspace{-2.8cm}
\centerline{\psfig{figure=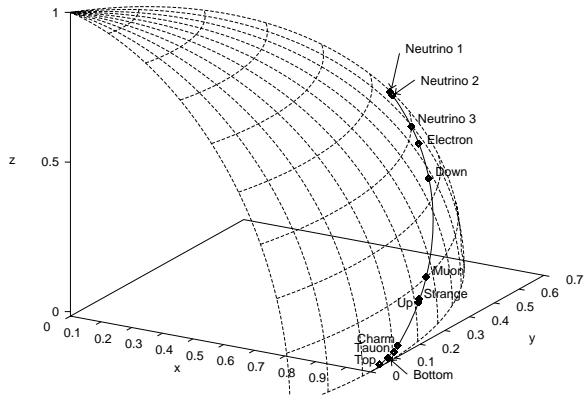,width=0.33\textwidth}}
\vspace{-.4cm} 
\caption{The trajectory of $(x',y',z')$ as the energy scale
varies.}
\label{trajectory}
\end{figure}

One interesting feature for the calculation outlined above is that the
trajectories traced out by the vector $(x', y', z')$ for the 4 different
fermion-types $U, D, L, N$ all coincide to a very good approximation,
only with the 12 physical fermion states at different locations 
(Figure \ref{trajectory}).
The points $(1,0,0)$ and $\frac{1}{\sqrt{3}}
(1,1,1)$ are fixed points so that the rate of flow is slower near the 
ends of the trajectory than in the middle.  For this reason, the states 
$t$ and $b$ are close together in spite of their big mass difference.
This observation will be 
of relevance later.

\begin{figure}[htb]
\vspace{-2.8cm}
\centerline{\psfig{figure=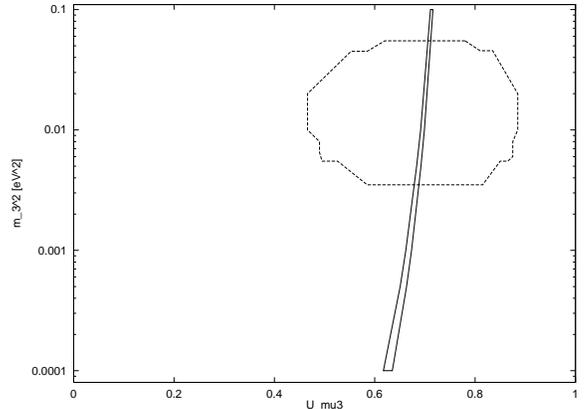,width=0.3\textwidth}}
\vspace{-.4cm}
\caption{90 \% CL limits on $U_{\mu 3}$ compared with DSM calculation.}
\label{Umu3}
\end{figure}

Since neutrino oscillations \cite{SuperK} are of particular interest 
at this conference,
let us take a closer look \cite{Bordesetal2}.
The element $U_{\mu3}$ of the lepton CKM matrix giving the mixing between 
the muon neutrino $\nu_\mu$ and the heaviest neutrino $\nu_3$ is constrained 
mainly by the data on atmospheric neutrinos.  From the old Kamiokande data 
\cite{Kamiokande} an analysis by Giunti et al.\cite{Giunti} gives the
bounds on $U_{\mu3}$ shown in Figure \ref{Umu3}.  In the DSM scheme, with 
parameters already fixed by the fit to the quark sector \cite{Bordesetal1}, 
the elements of $|U_{rs}|$ are calculable given the masses of 
$\nu_3$ and $\nu_2$.  Then, with $m_{\nu_2}$
taken in the range $10^{-11} {\rm eV}^2 < m_2^2 < 10^{-10} {\rm eV}^2$
as suggested by the Long Wave-Length Osicillation (LWO) (or the `vacuum'
or `just-so') solution to the solar neutrino problem \cite{Barger,Krastev},
the predicted band of values of $|U_{\mu3}|$ for a range of input values of 
$m_{\nu_3}$ is shown in Figure \ref{Umu3},
passing right through the middle of the allowed region.  No similar 
detailed analysis of the new SuperKamiokande data \cite{SuperK} has 
yet been performed, but the predicted band can be seen to
remain well within the allowed region: $.53 < U_{\mu 3} 
< .85,\ 5 \times 10^{-4} < m^2_3 < 6 \times 10^{-3} {\rm eV}^2$.

\begin{figure}[htb]
\vspace{-2.8cm}
\centerline{\psfig{figure=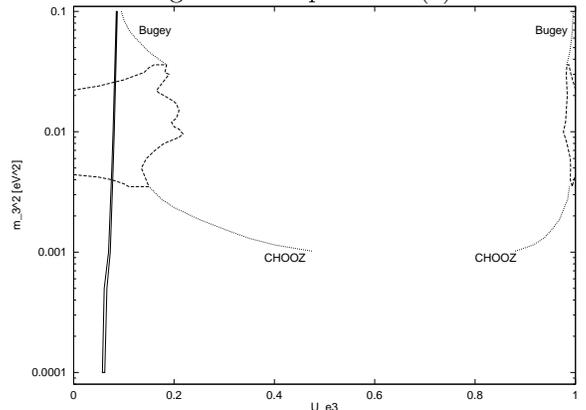,width=0.3\textwidth}}
\vspace{-.4cm}
\caption{90 \% CL limits on $U_{e 3}$ compared with DSM calculation.}
\label{Ue3}
\end{figure}

The same calculation gives the prediction shown in Figure \ref{Ue3} for 
the element $U_{e3}$ representing the mixing between the electron neutrino 
$\nu_e$ and $\nu_3$, which is constrained mainly by the reactor data 
from CHOOZ \cite{Chooz} and Bugey \cite{Bugey}.  If $m_3^2$ is higher than
$2 \times 10^{-3} {\rm eV}^2$, as favoured by the old Kamiokande data 
\cite{Kamiokande,Giunti} and the new data from Soudan reported in this
conference \cite{Soudan}, then the negative result from CHOOZ restricts 
$U_{e3}$ to quite small values, as indicated in Figure \ref{Ue3} and 
quoted in (\ref{expckml}).  The new SuperKamiokande data \cite{SuperK}
gives a best-fit value for $m_3^2$ of $2.2 \times 10^{-3}$, still implying
by CHOOZ a small value for $U_{e3}$, but do not exclude 
lower values of $m_3$ and hence
much larger values of $U_{e3}$.  In any 
case, as seen in Figure \ref{Ue3}, the band of values predicted by the 
DSM calculation falls always comfortably within the allowed region.

The DSM results summarized above for 
neutrino oscillations were obtained assuming 
$m_2^2$ of order $10^{-11}$ to $10^{-10} {\rm eV}^2$, as suggested
by the LWO solution \cite{Barger,Krastev}.
The alternative MSW \cite{MSW} solutions for solar neutrinos 
require \cite{MSWfit} $m_2^2 \sim
10^{-5} {\rm eV}^2$, for which no sensible DSM 
solution was found \cite{Bordesetal2}. 
It is thus intriguing to hear in this conference that the new SuperKamiokande 
data on the day-night variation and energy spectrum of solar neutrinos 
\cite{SuperK} also favours the LWO solution.

Further, generation being identified with dual colour in DSM, one expects
only 3 generations of neutrinos.
Thus, the result from
Karmen \cite{Karmen} reported in this conference against the existence of 
another neutrino with mass of order eV, as previously suggested by the 
LSND experiment \cite{LSND}, is also in the DSM's favour.

It is particularly instructive to compare the CKM matrices for leptons 
and quarks.  Both the empirical (\ref{expckmq}), (\ref{expckml}) 
and the calculated (\ref{calckmq}), (\ref{calckml}) matrices 
show the following salient features: \\
\hspace*{.5cm}$\bullet$ that the 23 element for leptons is much larger
than\\ 
\hspace*{.8cm}that for quarks,\\ 
\hspace*{.5cm}$\bullet$ that the 13 elements for both quarks and
leptons\\ 
\hspace*{.8cm}are much smaller than the rest, \\
\hspace*{.5cm}$\bullet$ that the 12 element is largish and comparable 
in\\ 
\hspace*{.8cm}magnitude for quarks and leptons. \\ 
These features, all so crucial for interpreting existing 
data, not only are all correctly reproduced by DSM 
calculation, but also can be understood within the 
scheme using some classical differential geometry 
as follows \cite{Bordesetal3}.

First, it turns out \cite{Chantsou1,Bordesetal1} 
that in the limit when the separation between the top 2 generations is small
on the trajectory traced out by $(x',y',z')$, which is the case for all 
4 fermion-types as seen in Figure \ref{trajectory}, then the vectors for 
the 3 generations form a Darboux triad \cite{Docarmo} composed of (i) the
radial vector $(x',y',z')$ for the heaviest generation, (ii) the tangent
vector to the trajectory for the second generation, and (iii) the vector
normal to both the above for the lightest generation.  The CKM matrix
is thus just the matrix which gives the relative orientation between 
the Darboux triads for the two fermion-types concerned.   
Secondly, by the Serret--Frenet--Darboux formulae, it follows
that the CKM matrix
can be written, to first order in the separation 
$\Delta s$ on the trajectory between $t$ and $b$ for quarks and 
between $\tau$ and $\nu_3$ for 
leptons, as
\begin{equation}
CKM \sim \left( \begin{array}{ccc} 1 & -\kappa_g \Delta s & -\tau_g \Delta s \\
                                   \kappa_g \Delta s & 1 & \kappa_n \Delta s \\
                                   \tau_g \Delta s & -\kappa_n \Delta s & 1 
   \end{array} \right),
\label{Darboux}
\end{equation}
where $\kappa_n$ and $\kappa_g$ are respectively the normal and geodesic
curvature and $\tau_g$ is the geodesic torsion of the trajectory.  Lastly, 
for our case of a curve on the unit sphere,  
$\kappa_n = 1$ and $\tau_g =0$, from which it follows that :\\
\hspace*{.5cm}$\bullet$ the 23 element equals roughly $\Delta s$,\\
\hspace*{.5cm}$\bullet$ the 13 element is of second order in $\Delta s$,\\
\hspace*{.5cm}$\bullet$ the 12 element depends on the details of the curve.\\
In Figure \ref{trajectory}, $\Delta s$ 
between $\tau$ and $\nu_3$ is much larger than that 
between $t$ and $b$, hence also the 23 element 
of the CKM matrix.  Indeed, measuring the actual separations in Figure 
\ref{trajectory}, one obtains already values very close to the actual CKM 
matrix elements in (\ref{calckmq}) and (\ref{calckml}) 
or in (\ref{expckmq}) and (\ref{expckml}).
The 13 elements should be small in both cases, as already noted.  As for
the 12 elements, they depend on both the locations and details of the curve, 
which explains why they need not differ much between quarks and leptons
in spite of the difference in separation, and also why the DSM prediction
in (\ref{calckml}) is not as successful for this element as for the others.

To test DSM further, one seeks
predictions outside the Standard
Model framework.  These are not hard to come by.  Identifying generation 
to dual colour, which is a local gauge symmetry, makes it imperative that 
any particle carrying a generation index can interact via the exchange of 
the dual colour gauge bosons, leading to flavour-changing 
neutral current (FCNC) effects. 
Given the calculations on the CKM matrices 
outlined above, all low energy FCNC effects can now be calculated in 
terms of a single mass parameter $\zeta$ related to the vev's of the dual 
colour Higgs fields \cite{Bordesetal4}.   
Inputting the mass difference $K_L - K_S$ which happens to give the tightest 
bound on $\zeta \sim 400 {\rm TeV}$, one obtains bounds on the
branching ratios of various FCNC decays.
In the following paragraph, an argument will be given which converts these 
bounds into actual order-of-magnitude estimates.    In
particular, the mode $K_L \rightarrow \mu^\pm e^\mp$ has a predicted 
branching ratio of around $10^{-13}$, 
less than 2 orders away
from the new BNL bound of  $5.1 \times 10^{-12}$ reported in this
conference \cite{BNL}. 

Since neutrinos carry a generation index, it follows that 
they will also acquire a new interaction through the exchange of 
dual colour bosons.  At low energy, this interaction will be very weak, 
being suppressed by the large mass parameter $\zeta$.  However, at C.M.
energy above $\zeta$, this new interaction will become strong. With
an estimate of 
at least $400 {\rm TeV}$, the predicted new
interaction is not observable in laboratory experiments at present or
in the foreseeable future, but it may be accessible in cosmic rays.  
For a neutrino colliding with a nucleon at rest in our atmosphere, 400 
TeV in the centre of mass corresponds to an incoming energy of about
$10^{20} {\rm eV}$.  Above this energy, neutrinos could thus in principle 
acquire a strong interaction and produce air
showers in the atmosphere.  Now air showers at and 
above this energy have been observed.  They have long been a 
puzzle to cosmic ray physicists since they cannot be due to proton or
nuclear primaries which would be quickly degraded from these energies by 
interaction with the 2.7 K microwave background \cite{Boratav}.  Indeed, 
the GZK cutoff \cite{Greisemin} for protons is
at around $5 \times 10^{19} 
{\rm eV}$.  Neutrinos, on the other hand, are not so affected by the
microwave background.
Hence, if they can indeed produce air showers via the new interaction
predicted by the DSM, they can give a very neat explanation for the old
puzzle of air showers beyond the GZK cut-off \cite{Bordesetal5}.  
Further tests for the proposal have been suggested \cite{Bordesetal6}.
The proposal also gives a
rough upper bound on the mass parameter $\zeta$ 
governing FCNC effects which is close to the lower bound obtained in the 
preceding paragraph.  It was on the basis of this coincidence that the 
above FCNC bounds were converted into 
actual order-of-magnitude estimates.

\begin{figure}[h]
\center
\begin{picture}(240,400)
\put(60,350){\framebox(120,18){\shortstack{{\small NONABELIAN DUALITY}\\
{\scriptsize{Chan--Faridani--Tsou \cite{Chanftsou1,Chanftsou2}}}}}}
\put(116,340){\line(1,0){8}}
\put(120,344){\line(0,-1){8}}
\put(0,320){\framebox(110,10){{\small STANDARD MODEL}}}
\put(130,320){\framebox(110,10){{\small 'T HOOFT THEOREM
\cite{thooft}}}}
\put(55,320){\line(0,-1){8}}
\put(185,320){\line(0,-1){8}}
\put(55,312){\line(1,0){130}}
\put(120,312){\vector(0,-1){12}}
\put(45,282){\framebox(150,18){\shortstack{{\small DUALIZED STANDARD MODEL}\\
{\scriptsize{Chan--Tsou \cite{Chantsou1}}}}}}
\put(120,282){\line(0,-1){10}}
\put(55,272){\line(1,0){130}}
\put(55,272){\vector(0,-1){12}}
\put(185,272){\vector(0,-1){12}}
\put(0,240){\framebox(110,20){\shortstack{{\small HIGGS FIELDS}\\
{\small (AS FRAME VECTORS)}}}}
\put(130,240){\framebox(110,20){\shortstack{{\small 3 GENERATIONS}\\
{\small (AS DUAL COLOUR)}}}}
\put(30,240){\vector(0,-1){18}}
\put(160,240){\line(0,-1){6}}
\put(80,234){\line(1,0){80}}
\put(80,234){\vector(0,-1){12}}
\put(0,202){\framebox(110,20){\shortstack{{\small SYMMETRY}\\
{\small BREAKING PATTERN}}}}
\put(55,202){\vector(0,-1){16}}
\put(0,146){\framebox(156,40){\shortstack{{\small FERMION MASS HIERARCHY:}\\
e.g. $m_t \gg m_c \gg m_u$\\{\small FERMION MASSES AND MIXINGS}\\
{\small CALCULABLE PERTURBATIVELY}}}}
\put(55,146){\vector(0,-1){16}}
\put(0,84){\framebox(156,46){\shortstack{{\small (1-LOOP, 1ST ORDER)}\\
{\small QUARK AND LEPTON MASSES}\\{\small QUARK CKM MATRIX}\\
{\scriptsize{Bordes--Chan--Faridani--Pfaudler--Tsou
\cite{Bordesetal1}}}}}}
\put(106,84){\line(0,-1){10}}
\put(106,74){\line(1,0){74}}
\put(180,74){\vector(0,-1){19}}
\put(55,84){\vector(0,-1){54}}
\put(0,0){\framebox(156,30){\shortstack{{\small (EXTENDED TO NEUTRINOS)}\\
{\small NEUTRINO OSCILLATIONS}\\
{\scriptsize{Bordes--Chan--Pfaudler--Tsou
\cite{Bordesetal2,Bordesetal3}}}}}}
\put(210,240){\vector(0,-1){24}}
\put(166,176){\framebox(74,40){\shortstack{{\small NEUTRINOS}\\
{\small INTERACT}\\{\small STRONGLY AT}\\ {\small HIGH ENERGY}}}}
\put(210,176){\vector(0,-1){16}}
\put(164,90){\framebox(76,70){\shortstack{{\small `EXPLANATION'}\\
{\small FOR PUZZLE OF}\\{\small AIR SHOWERS}\\ {\small AT
$E>10^{20}$eV}\\
{\scriptsize{Bordes--Chan--}}\\{\scriptsize{Faridani--Pfaudler--}}\\
{\scriptsize{Tsou \cite{Bordesetal5,Bordesetal6}}}}}}
\put(210,90){\vector(0,-1){35}}
\put(166,0){\framebox(74,55){\shortstack{{\small PREDICTIONS}\\
{\small OF B.R.\ FOR}\\{\small FCNC DECAYS}\\
{\scriptsize{Bordes--Chan--}}\\{\scriptsize{Faridani--Pfaudler--}}\\
{\scriptsize{Tsou \cite{Bordesetal4}}}}}}
\end{picture}
\caption{Summary flow-chart}
\label{Summary}
\end{figure}
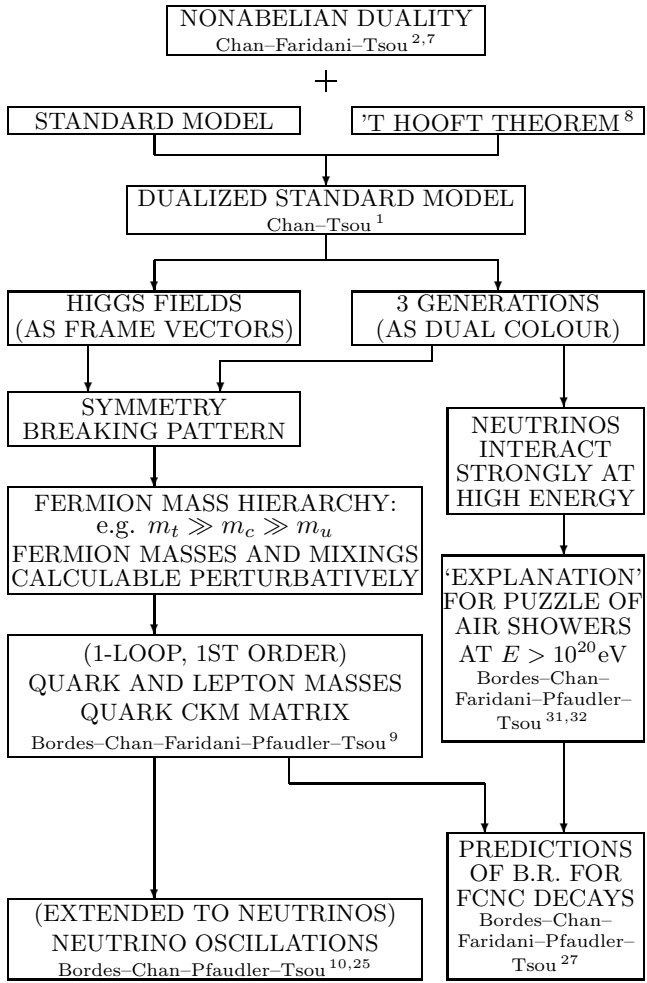

The conclusions are summarized in Figure \ref{Summary}.

It is a pleasure for us to acknowledge our profitable and most enjoyable
collaboration with Jacqueline Faridani and Jakov Pfaudler. TST also 
thanks the Royal Society for a
travel grant to Vancouver.

\end{document}